\documentclass[useAMS,usenatbib]{mn2e}
\usepackage{amssymb,amsmath,epsfig,times, natbib,color}
\pdfoutput=1

\title[Bays in clusters]{Is there a giant Kelvin-Helmholtz instability in the sloshing cold front of the Perseus cluster?}
\author[S. A. Walker et al] {\parbox[]{6.5in}{{
      S. A. Walker,$^1$\thanks{Email: 
    stephen.a.walker@nasa.gov}, J. Hlavacek-Larrondo$^2$, 
    M. Gendron-Marsolais$^2$, A. C. Fabian$^3$,
     H. Intema$^4$, J. S. Sanders$^5$, J. T. Bamford$^6$ and
    R. van Weeren$^7$ 
    }\\
     \footnotesize
     $^1$Astrophysics Science Division, X-ray Astrophysics Laboratory, Code 662, NASA Goddard Space Flight Center, Greenbelt, MD 20771, USA \\
$^2$Departement de Physique, Universite de Montreal, Montreal, QC H3C 3J7,
Canada \\
       $^3$Institute of Astronomy, Madingley Road, Cambridge CB3 0HA \\
      $^4$ Leiden Observatory, Leiden University, Niels Bohrweg 2, NL-2333CA,
Leiden, The Netherlands \\
  $^5$Max-Planck-Institute fur extraterrestrische Physik, 85748 Garching, 
Germany  \\
$^6$Department of Applied Mathematics, The University of Leeds, Leeds LS2 9JT \\
$^7$ Harvard-Smithsonian Center for Astrophysics, 60 Garden Street, Cambridge, MA 02138, USA \\
    }}

\date{}

\begin{document}

\maketitle

\begin{abstract}
Deep observations of nearby galaxy clusters with Chandra have revealed concave
`bay' structures in a number of systems (Perseus, Centaurus and Abell 1795),
which have similar X-ray and radio properties. These bays have all the properties of cold fronts, where the temperature rises and density falls sharply, but are 
concave rather than convex. By comparing to simulations of gas
sloshing, we find that the bay in the Perseus cluster bears a striking
resemblance in its size, location and thermal structure, to a giant ($\approx$50 kpc) roll resulting
from Kelvin-Helmholtz instabilities. If true, the morphology of this structure
can be compared to simulations to put constraints on the initial average ratio of the thermal
and magnetic pressure, $\beta= p_{\rm th} / p_{\rm B}$, throughout the overall cluster before the sloshing occurs, for which we find
$\beta=200$ to best match the observations. Simulations with a stronger magnetic field ($\beta=100$) are disfavoured, as in these the large Kelvin-Helmholtz rolls do not form,
while in simulations with a lower magnetic field ($\beta=500$) the level of instabilities is much larger than is observed. We find that the bay structures in Centaurus and Abell 1795 may also be explained by such features of gas sloshing.
\end{abstract}


\section{Introduction}
Chandra observations of the cores of nearby relaxed galaxy clusters have
revealed a panoply of structures in the intracluster medium (ICM). Active
Galactic Nuclei (AGN) are seen to inflate bubbles which expand and rise
outwards (\citealt{Fabian2000}, \citealt{McNamara2000}, \citealt{Fabian2012}).
Minor mergers are seen to induce gas sloshing of the cool core,
resulting in spiral patterns of sharp cold fronts, interfaces where the
temperature and density jumps dramatically on scales much smaller than the mean
free path (\citealt{Markevitch2000}, \citealt{Markevitch2007}). The resulting
imprints of these processes in the ICM provide powerful
tools for unravelling both the physics of the ICM itself, and the AGN feedback
believed to be responsible for preventing runaway cooling. 

\begin{figure*}
  \begin{center}
    \leavevmode
\includegraphics[width=0.9\linewidth]{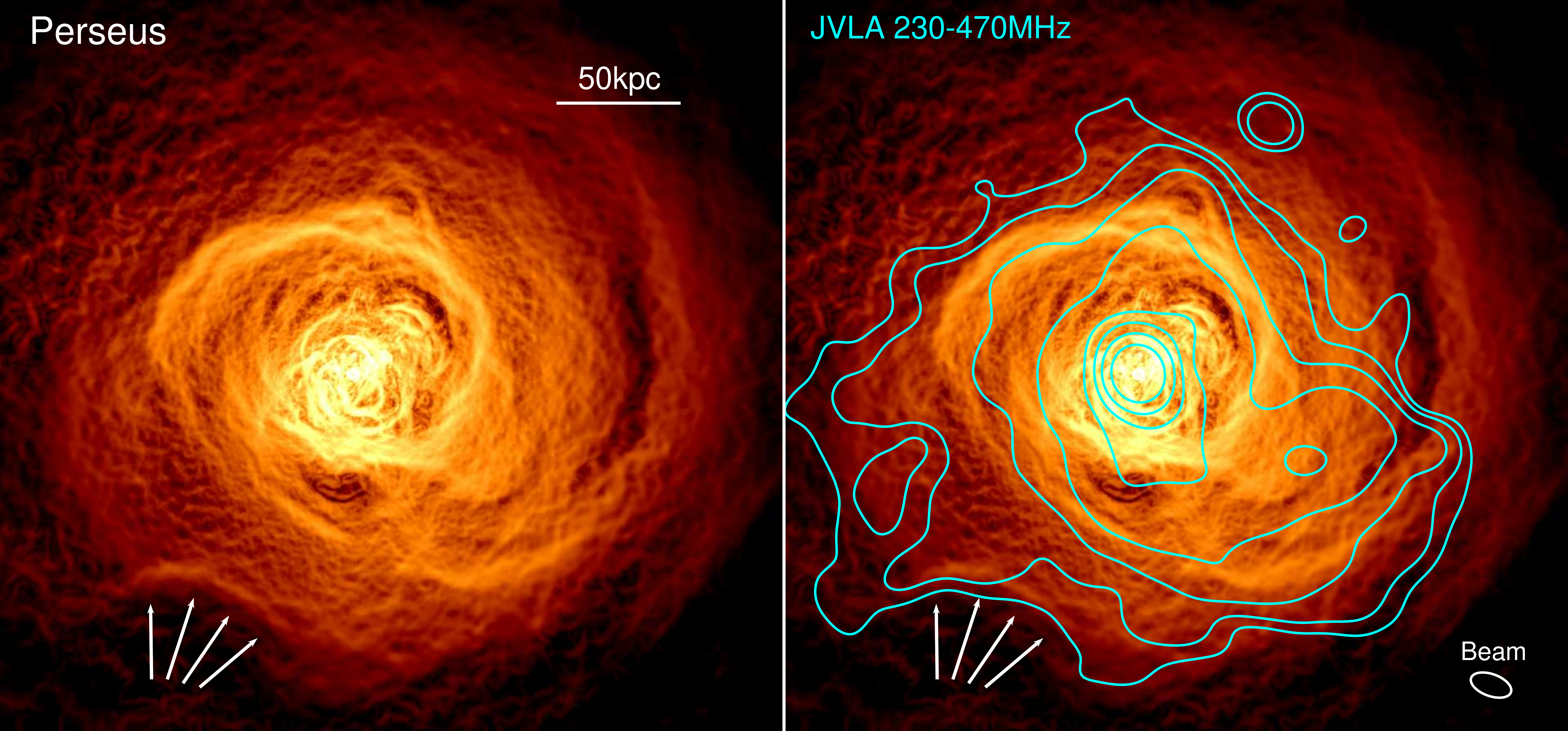},
	\includegraphics[width=0.9\linewidth]{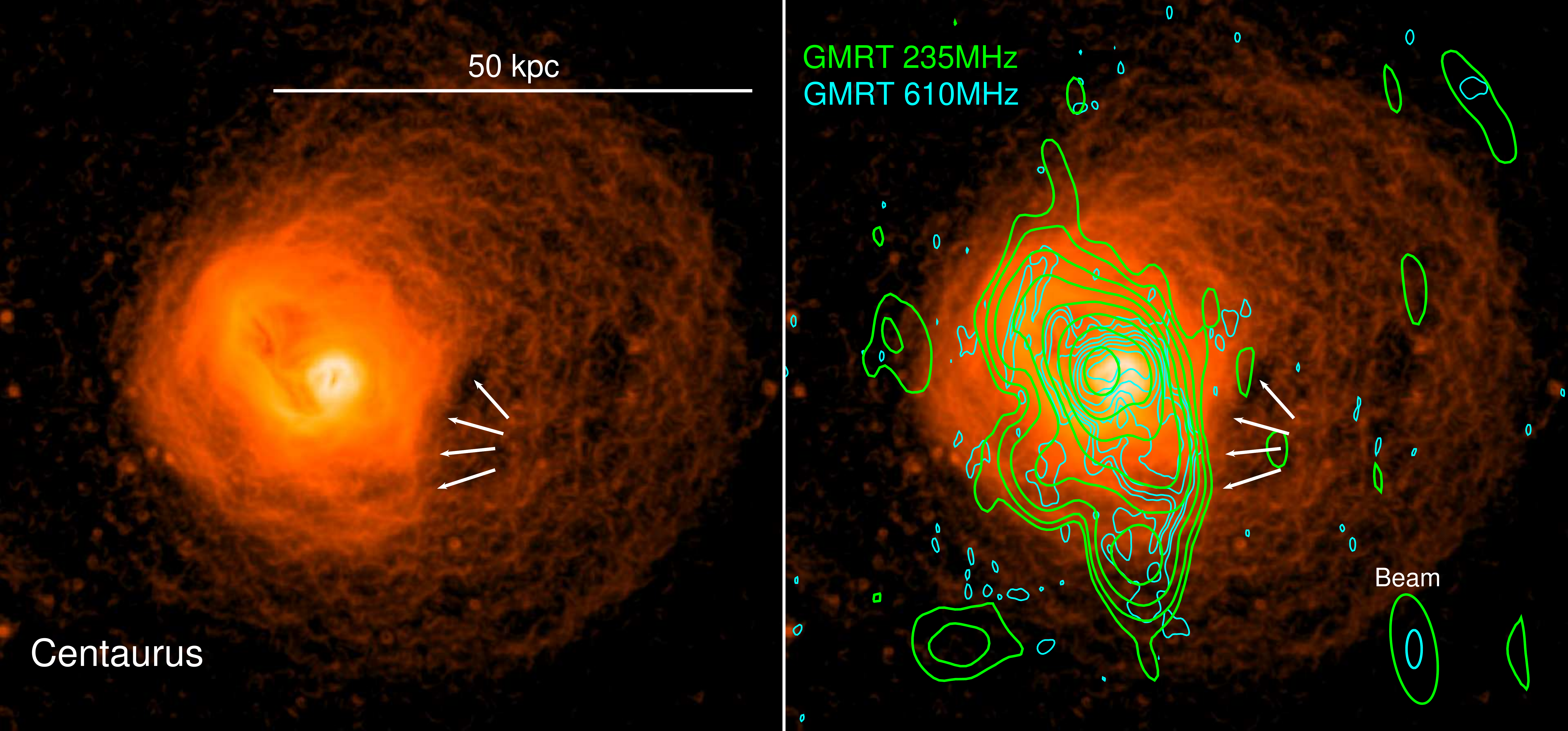},
\includegraphics[width=0.9\linewidth]{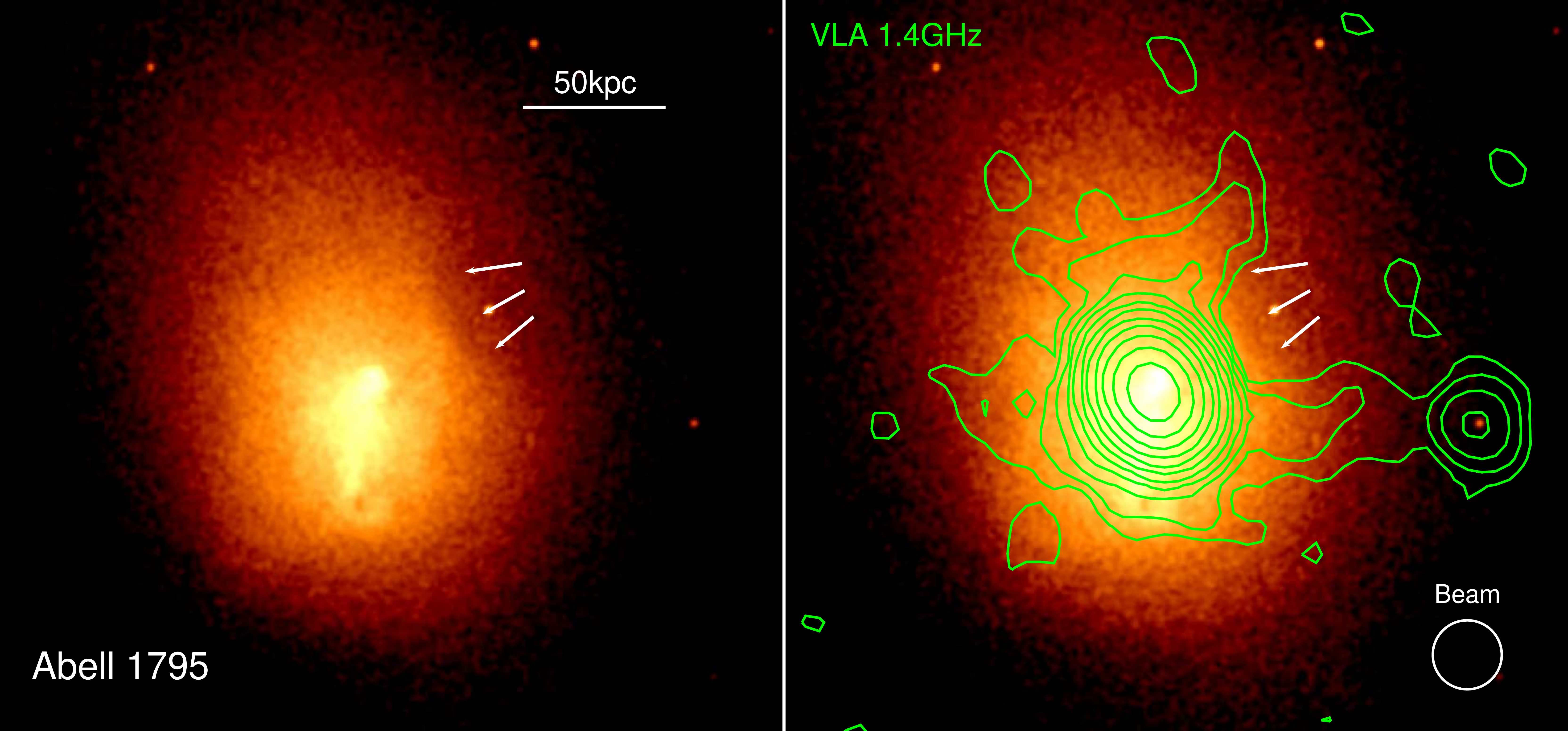},

      \caption{Comparing the bays in the Perseus cluster (top), Centaurus
(middle) and A1795 (bottom). Radio contours are overplotted on the X-ray data in the right hand panels.
The radio emission is constrained behind the bays, which themselves contain no
radio emission, the opposite to what would be expected for bubbles inflated by
AGN feedback. For Perseus and Centaurus we have filtered the images with the GGM filter to emphasise gradient structure.}
      \label{compare_xraywithradio}
  \end{center}
\end{figure*}

In at least three nearby relaxed clusters (Perseus, Centaurus and A1795), these structures
include
unusual concave `bay'-like features, which are not easily explained by either AGN
feedback or gas sloshing (see \citealt{Fabian2006}, \citealt{Sanders2016} and
\citealt{Walker2014_A1795}), and these are marked by the white arrows in Fig. \ref{compare_xraywithradio}.

 These sharp surface brightness discontinuities have
all the properties of cold fronts, namely a temperature increase from the more
dense side to the less dense side, and widths which are of the same order as the
Coulomb mean free path. However they have a concave curvature, which contrasts with the standard convex curvature of sloshing cold fronts,
which has led such features to be also interpreted as the inner rims of
cavities. Here we investigate the possible formation scenarios for these
`bays', comparing their properties in different clusters using a multiwavelength approach of deep Chandra
and radio observations, together with simulations of gas sloshing (\citealt{ZuHone2016}) and cavities. 

Whilst appearing similar in X-ray images, the inner rims of AGN inflated bubbles should have different radio properties to 
inverted cold fronts. Radio mini-haloes in clusters tend to be confined behind cold fronts, with a sharp drop in radio 
emission across the cold front edge (\citealt{Mazzotta2008}). AGN inflated bubbles on the other hand should be filled with 
radio emitting relativistic plasma. A multiwavelength approach will therefore allow us to break this degeneracy.     

Simulations of gas sloshing in relaxed clusters predict that as cold fronts rise
outwards and age, Kelvin-Helmholtz instabilities (KHI), brought about by the velocity
shear between the cool sloshing core and the outer, hotter cluster ICM can form (e.g. \citealt{ZuHone2011}, \citealt{Roediger2012}, \citealt{Roediger2013}).
In simulations, these can grow to sizes of the order of tens of kpc for old cold
fronts. These Kelvin Helmholtz rolls can produce inverted cold fronts, which are concave, similar to the `bays' we observe.

In simulations, the development of KHI rolls is very sensitive to the strength of the magnetic field and the level of viscosity, 
with vastly different structures forming depending on the input values for these. Differences in the cluster microphysics 
can therefore affect the cold front morphology on scales of tens of kpc (for a review see \citealt{Zuhone2016review}).
Because of this, observing large KHIs in real clusters would provide powerful constraints on the 
magnetic field strength and viscosity in the cluster ICM.  

In section \ref{sec:data} the X-ray and radio data used are discussed. Section \ref{sec:bayproperties} compares
the properties of the bays in the Perseus, Centaurus and Abell 1795. In sections \ref{sec:cavsim} and \ref{sec:sloshsim} we 
compare our observations to simulations of cavities and gas sloshing, respectively. In section \ref{sec:conclusions} we present our conclusions. We use a standard $\Lambda$CDM cosmology with $H_{0}=70$  km s$^{-1}$
Mpc$^{-1}$, $\Omega_{M}=0.3$, $\Omega_{\Lambda}$=0.7. All errors unless
otherwise stated are at the 1 $\sigma$ level. 

In this work, the term `bay' refers to the concave surface brightness discontinuity itself, which, in the analogy with 
an ocean bay, more accurately corresponds to the `shoreline' between the water and the land. We refer to the side of the bay towards the 
cluster center as `behind' the bay, while the opposite side is `in front' of the bay.  

\section{Data}
\label{sec:data}
\subsection{X-ray data}
We use deep Chandra observations of Perseus (900ks of ACIS-S data, plus 500ks of ACIS-I wide field observations), 
Abell 1795 (710ks of ACIS-S and ACIS-I) and the
Centaurus cluster (760ks of ACIS-S), tabulated in table \ref{obsdata} in Appendix \ref{appendix_obs}. The data used, and the reduction process, are described in \citet{Fabian2006} and 
\citet{Fabian2011} for Perseus, in \citet{Walker2014_A1795} for Abell 1795, and in \citet{Sanders2016} and \citet{Walker2015c}
for Centaurus. 

In short, the Chandra data were reduced using the latest version of CIAO (4.8). The events were reprocessed using 
\textsc{chandra\_repro}. Light curves were then extracted for each observation, and periods of flaring were removed. 
Stacked images in the broad 0.7-7.0 keV band were created by first running the script \textsc{reproject\_obs} to 
reproject the events files, and then \textsc{flux\_obs} was used to extract images and produce their exposure maps. 
For each cluster, the observations were stacked, weighting by the exposure map. 
  
Spectra were extracted using \textsc{dmextract}, with ARFs and RMFs created using \textsc{mkwarf} and 
\textsc{mkacisrmf}. The script \textsc{acis\_bkgrnd\_lookup} was used to find appropriate blank sky background 
fields for each observation, which were rescaled so that their count rates in the hard 10-12keV band matched 
the observations.

\subsection{Radio data}

We use deep Karl G. Jansky Very Large Array (JVLA) observations of
Perseus (contours shown in the top right panel of Fig. \ref{compare_xraywithradio}), consisting of 5 h in the B configuration (maximum antenna
separation of 11.1km, synthesized beamwidth of 18.5 arcsec) in the P-band
(230-470 MHz) obtained from a shared-risk proposal (2013 Hlavacek et al.).
The JVLA is outfitted with new broadband low frequency receivers with
wider bandwidth. The data reduction was performed with CASA (Common
Astronomy Software Applications). A pipeline has been specifically
developed to reduce this dataset and is presented in detail in
\citet{Gendron-Marsolais2017}.

The main steps of data reduction can be
summarised as follows. The RFI were identified both manually and
automatically. The calibration of the dataset was conducted after the
removal of most of the RFI, and each calibration table was visually
inspected, the outliers solutions were identified and removed. Parameters
of the clean task were carefully adjusted to take account of the
complexity of the structures of Perseus core emission and its high dynamic
range. We used a multi-scale and multi-frequency synthesis-imaging
algorithm, a number of Taylor coefficients greater than one, W-projection
corrections, a multi-scale cleaning algorithm and a cleaning mask limiting
regions where emission was expected. A self-calibration was also
performed, using gain amplitudes and phases corrections from data to
refine the calibration. The resulting image has an rms noise of 0.38
mJy/beam, a beam size of 22.0 $\times$ 11.4 arcsec and a maximum of 10.58
jy/beam.

The Giant Metrewave Radio Telescope \citep[GMRT;][]{1991ASPC...19..376S} was used to observe the center
 of the Centaurus cluster (NGC4696) during two 5-hour observe sessions in March 2012 (project 21\_006; PI
 Hlavacek-Larrondo). These contours are shown in the middle right panel of Fig. \ref{compare_xraywithradio}. Data were recorded simultaneously in single-polarisation mode at 235 and 610 MHz over
 16 and 32 MHz bandwidth, respectively, using 0.13 MHz frequency channel resolution and 16.1 second time resolution.
 The flux calibrator 3C\,286 was observed for 10--15 minutes at the start and end of both observe sessions.
 In between, the target field (NGC4696) was observed in scans of 30 minutes, interleaved with phase calibrator
 scans of 5 minutes. The total time on target is close to 7 hours.

The observational data at both frequencies were processed using the SPAM pipeline \citep{2014ascl.soft08006I}
 in its default mode. We started by processing of the 235 MHz data using a skymodel for calibration purposes
 derived from the GMRT 150 MHz sky survey \citep[TGSS ADR1;][]{2016arXiv160304368I}. The resulting 235 MHz image
 has a central image sensitivity of 0.95 mJy/beam and a resolution of $27.7'' \times 11.0''$ (PA 10 degrees). 
We used the PyBDSM source extractor \citep{2015ascl.soft02007M} to obtain a source model of the image, which
 we used as a calibration skymodel for processing of the 610 MHz data. The resulting 610 MHz image has a central 
image sensitivity of 85 $\mu$Jy/beam and a resolution of $9.5'' \times 3.8''$ (PA 0 degrees). In both images, 
the bright central radio source is surrounded by some image background artefacts due to dynamic range limitations
 that are known to exist for GMRT observations, increasing the local background rms by a factor of 2--3. However, 
the artefacts have little effect on the observed radio emission presented in this study.

The 1.4 GHz VLA contours for Abell 1795 shown in the bottom panel of Fig. \ref{compare_xraywithradio} are taken from 
\citet{Giacintucci2014}.

\begin{figure}
  \begin{center}
    \leavevmode
\includegraphics[width=\linewidth]{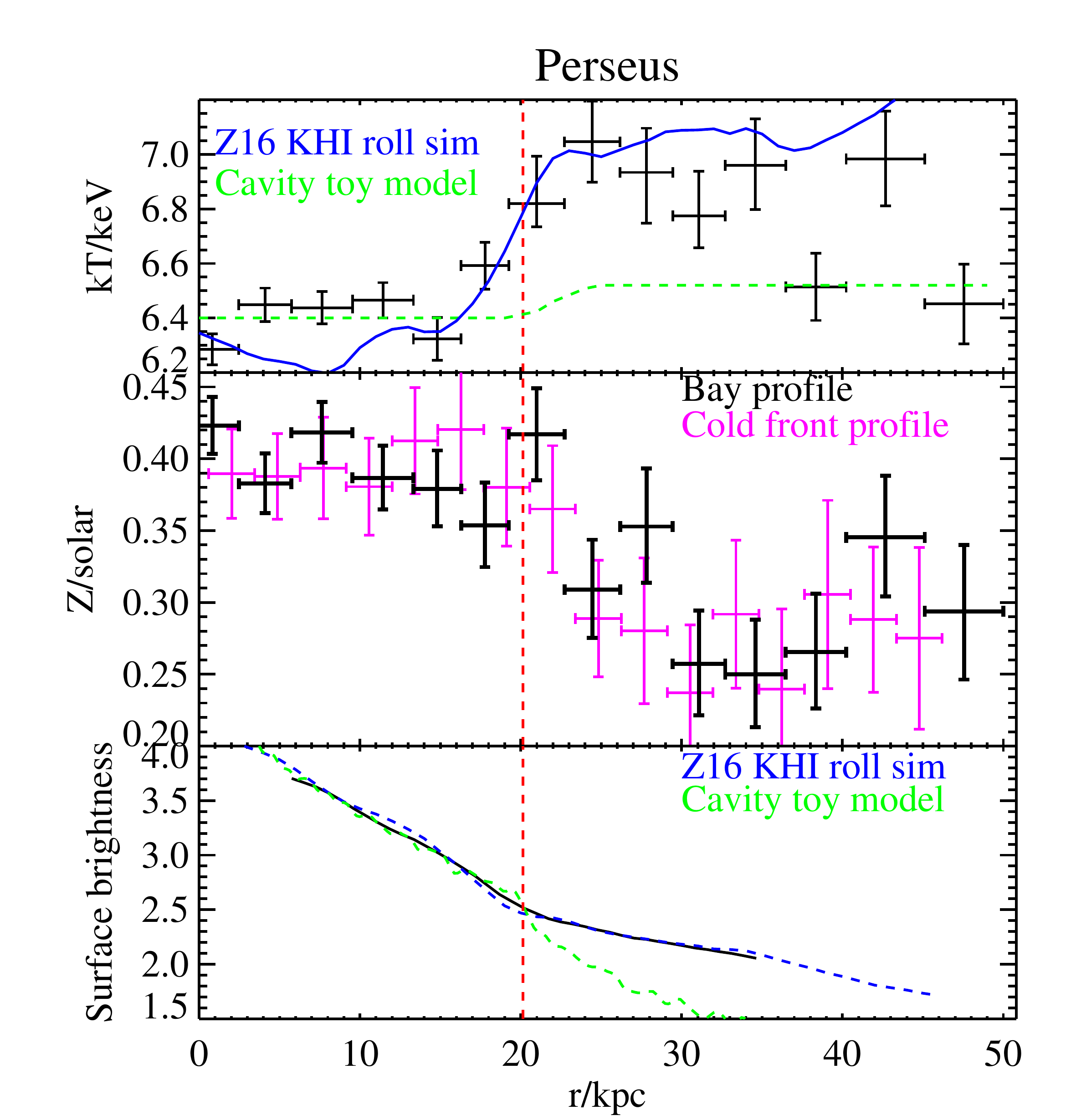}
      \caption{Temperature (top), metallicity (middle) and surface brightness
(bottom) profiles over the bay in Perseus. The location of the surface
brightness edge is shown by the dashed red line. The
temperature jump is clear. The fractional increase in
temperature resembles that seen across the bay shaped KH roll in the simulations
of \citet{ZuHone2016}
(Z16, blue line) shown later in Fig. \ref{perseus_comparetobeta200}, and is much larger than the temperature jump expected across an empty cavity in our cavity toy model (dashed green line).
 The metallicity also seems to sharply decrease across the edge (black points), in the same 
manner as the metal abundance drop over the `normal' part of the cold front next to the bay (magenta points). In
the bottom panel we compare
the projected surface brightness profile across the bay (black) with that across
the simulated KH roll (blue dashed) and a toy model of an empty spherical cavity in Perseus
(dashed green). The simulated KH roll profile agrees well with the magnitude of the observed
surface brightness drop, while the cavity toy model predicts a much bigger
surface brightness drop than is observed. As discussed in section \ref{sec:cavsim}, an ellipsoidal cavity toy model was also tested, but this increases the surface brightness profile discrepancy. }
      \label{Temperatures_finebayedge}
  \end{center}
\end{figure}

\begin{figure}
  \begin{center}
    \leavevmode
    \vbox{
\includegraphics[width=\linewidth]{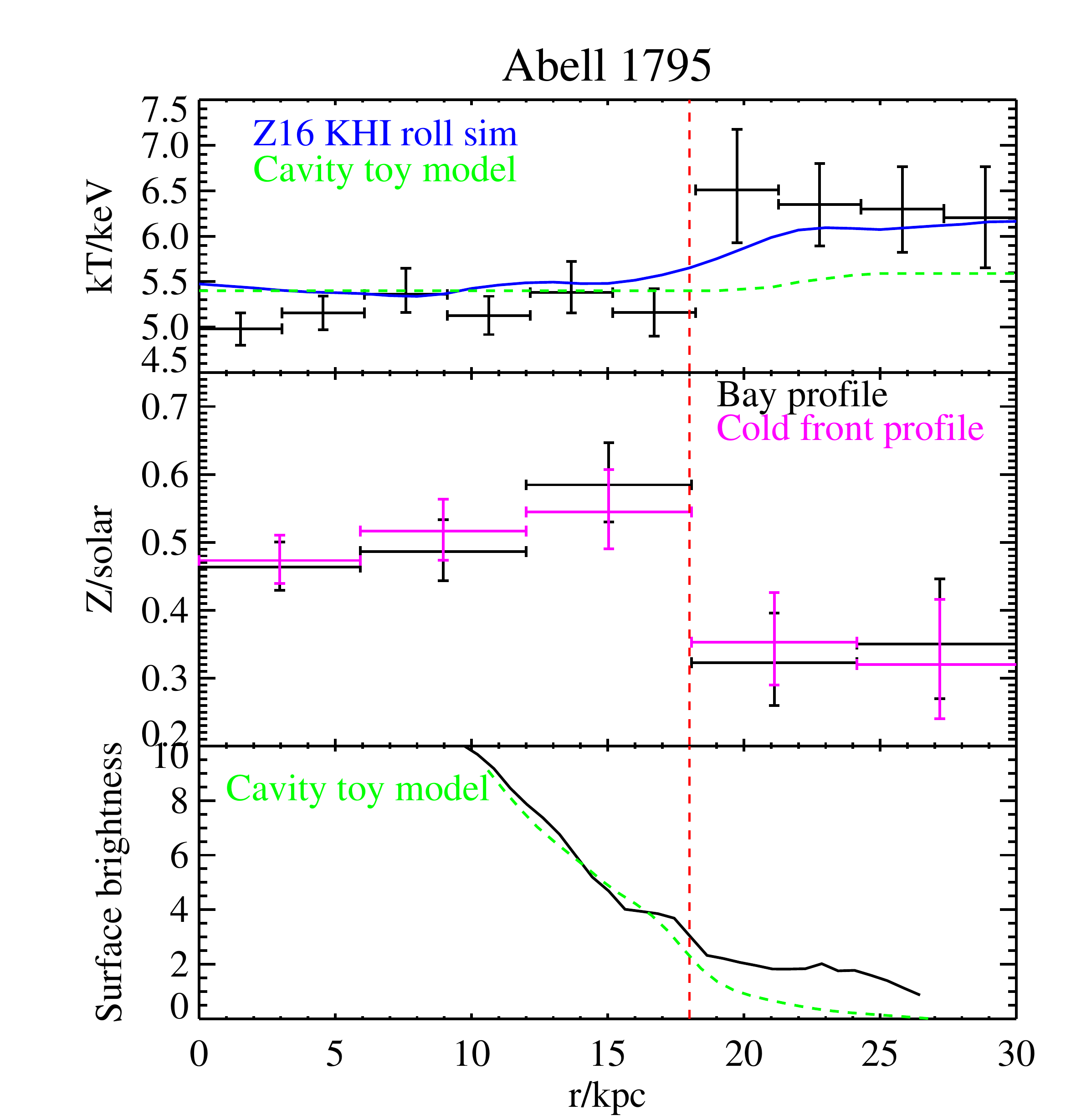}
\includegraphics[width=\linewidth]{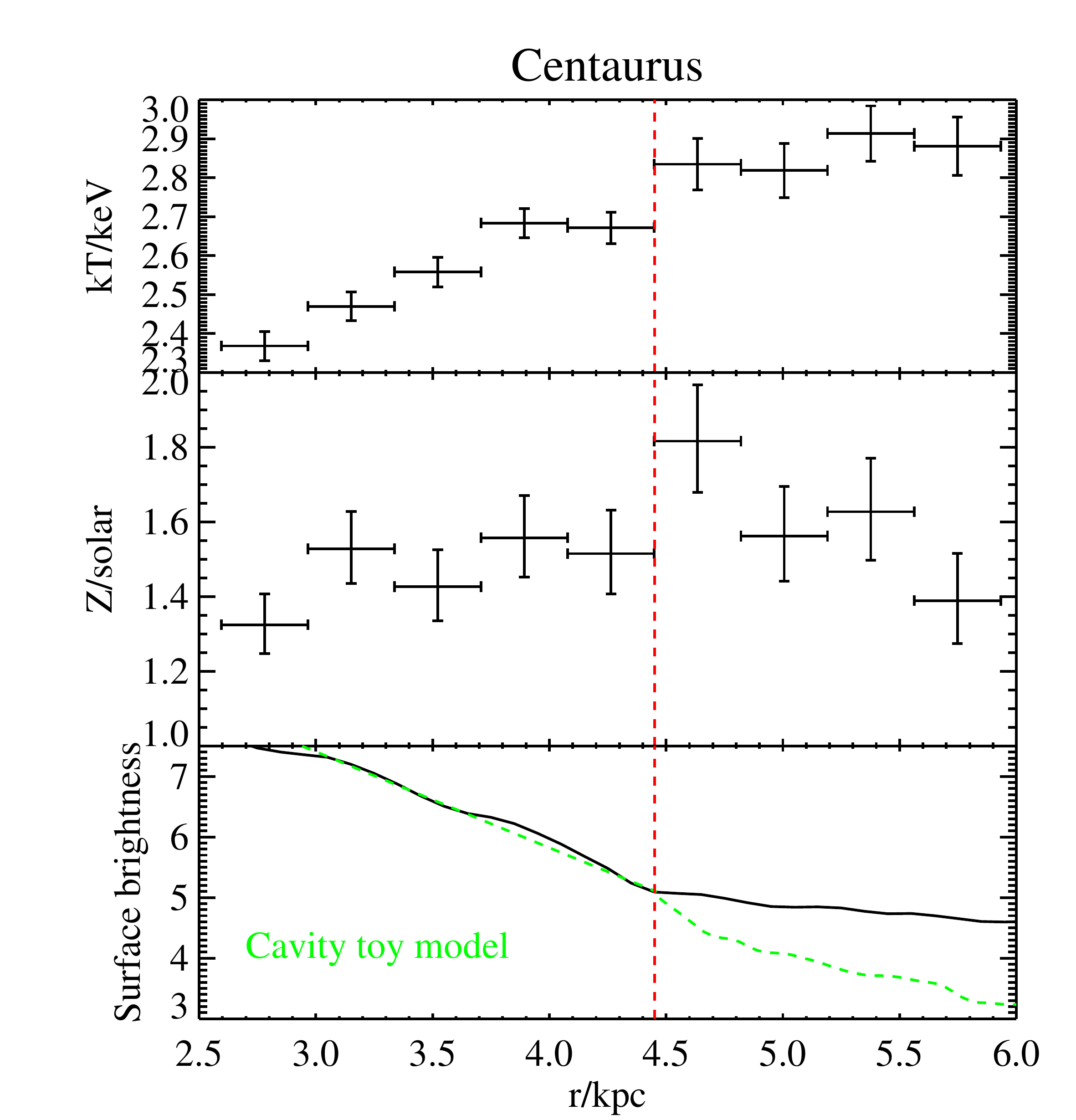}
}
      \caption{Same as Fig. \ref{Temperatures_finebayedge} but for the bays in Abell 1795
(top) and the Centaurus cluster (bottom). As with the Perseus bay, we see that the for Abell 1795 
the temperature jump is roughly consistent with that across a simulated KHI roll, and is much larger than the jump 
predicted by our empty cavity toy model. The cavity toy model again overestimates the surface brightness drop.   }
      \label{Temperatures_finebayedgeA1795andCen}
  \end{center}
\end{figure}

\begin{figure}
  \begin{center}
    \leavevmode
\includegraphics[width=\linewidth]{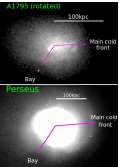}
      \caption{Here we have flipped and rotated the image of Abell 1795 to compare to Perseus. The separation angle 
between the main cold front and the bay in each system is similar at around 130 degrees.  }
      \label{Compare_Perseus_A1795rot}
  \end{center}
\end{figure}

\section{Bay properties}
\label{sec:bayproperties}
Figure \ref{compare_xraywithradio} shows Gaussian Gradient
Magnitude (GGM) filtered broad band (0.7-7.0keV) Chandra
X-ray images of Perseus (top) and Centaurus (middle). The GGM filter enhances surface
brightness edges in these images (\citealt{Sanders2016b}, \citealt{Walker2016}), increasing the contrast of the
edges by around a factor of 10 for Perseus and Centaurus. Due to the higher redshift and smaller angular size of A1795 (which is at z=0.062, compared to z=0.01 for Centaurus and z=0.018 for Perseus), the normal broad band Chandra
image is shown. An unfiltered version of the Perseus Chandra image is shown in the top left panel of Fig. \ref{perseus_comparetobeta200}, while an unfiltered version of the
Centaurus image is shown in figure 1 of \citet{Sanders2016}. The bays in each cluster are clear and marked
with the white arrows. In the right hand column we show the radio contours
overlain on the X-ray data. 

\subsection{Widths of the edges}

To determine the widths of the bays edges, we fit their surface brightness profiles with a broken powerlaw
model, which is convolved with a Gaussian, as in \citet{Sanders2016b} and \citet{Walker2016}. In all three cases we obtain
widths consistent with the known cold fronts in these clusters. For Perseus the upper limit on the width is 2 times the Coulomb mean 
free path, while in Centaurus the width of 2kpc is consistent with the range of widths (0-4kpc) found for the main cold front in \citealt{Sanders2016b}. In Abell 1795 the width is consistent with that of the main cold front to the south 
studied by \citet{Markevitch2001} and \citet{Ehlert2015}. This all indicates that transport processes are heavily suppressed across these edges, to the same extent as the known cold fronts. 

\subsection{Temperature, density and metallicity profiles}

In Figs. \ref{Temperatures_finebayedge} and
\ref{Temperatures_finebayedgeA1795andCen} we show projected temperature,
metallicity and surface brightness profiles across the edges of the bays in the
three clusters. The regions used for extracting these spectra are shown in Fig. \ref{Cluster_profile_regions} in Appendix \ref{appendix_obs}. In each case we see an abrupt temperature jump across the bay,
consistent with cold front behaviour. In Perseus and A1795 we also see a
significant decline in the metal abundance, from $\sim0.5$Z$_{\odot}$ to
$\sim0.3$Z$_{\odot}$, which is again consistent with cold front behaviour, where
the metal enriched cluster core is sloshing, leading to sharp falls in
metallicity across cold fronts (\citealt{Roediger2011}). This metallicity structure is also at odds with an AGN inflated bubble origin
for the bays, as typically the bubbles along the jet direction are found to
coincide with a metal enhancement (\citealt{Kirkpatrick2011}), as they uplift
metal enriched gas from the cluster core. When we continue the profiles further outwards from the cluster core, we see no increase in the 
metal abundance in front of the bay (i.e. no metal excess in what would be the middle of the cavity). When we compare the metal abundance profile across the bay with that across the `normal' parts of the cold
front next to the bay (which have a convex curvature) in the middle panels of Fig. \ref{Temperatures_finebayedge} and Fig. \ref {Temperatures_finebayedgeA1795andCen} for Perseus and Abell 1795 respectively 
(magenta points), we see that the two profiles are consistent with each other. 

\subsection{Radio properties}

All three clusters have radio mini-haloes in their cores. These radio
mini-haloes are relatively rare, and are confined to the central cooling region
of the clusters. Their origin remains a subject of continued debate. There are
two leading theories for the radio emission: one is that  
gas sloshing induced turbulence re-accelerates relativistic electrons in cluster
cores (originating from AGN feedback) (\citealt{Gitti2002},
\citealt{Gitti2004}), the second is that relativistic cosmic-ray protons
inelastically collide with thermal protons, generating secondary particles
(\citealt{Pfrommer2004}, \citealt{KeshetLoeb2010}, \citealt{Keshet2010}). Their
spatial extent is typically bound by cold fronts (\citealt{Mazzotta2008}),
believed to be the result of the draped magnetic fields around cold fronts
preventing the relativistic electrons from passing through them, constraining
them to the inside of the cold front (see the simulation work of
\citealt{ZuHone2013}). In Perseus, we see that the mini halo is constrained
behind the prominent cold front to the west, while in Centaurus and A1795 the
mini haloes are confined behind the cold fronts to the east and south
respectively.

Interestingly, we see that in all three clusters, the radio haloes are also
constrained behind the bays, adding support to the idea that these bays are cold
fronts which are concave rather than convex. Previously for
 Perseus, \citet{Fabian2011} compared the X-ray image to early 49cm 
VLA data from \citet{Sijbring1993} and reported that the bay is coincident with a minimum in radio flux. This radio behaviour is the
opposite to what would be expected from a bubble inflated through AGN feedback,
which are typically found to be filled with radio emitting relativistic plasma. For each cluster, the radio level in front of the bays is consistent with the background level. The typical
radio flux expected if these were cavities is at least the same order as that of the radio halo, and would be easily seen if present.

 \begin{figure*}
  \begin{center}
    \leavevmode
\includegraphics[width=\linewidth]{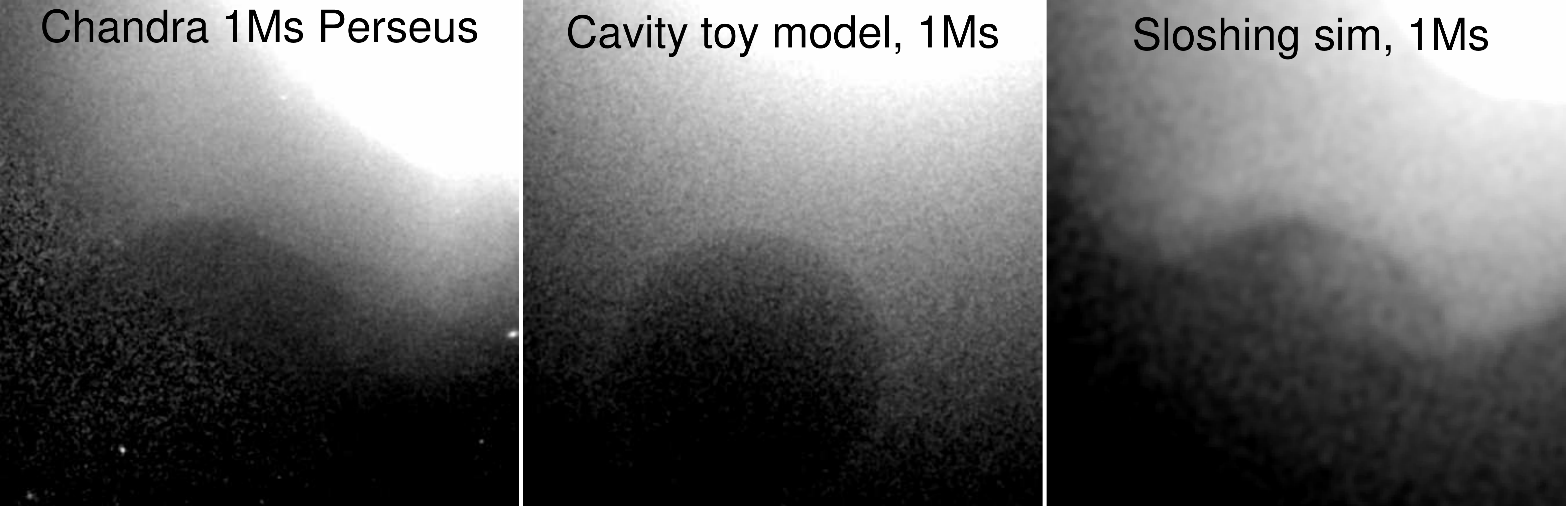}
      \caption{Comparing the image of the bay in Perseus (left) with a
simulation of an empty spherical cavity in the same location and with the same
size (centre), and the KH roll from the sloshing simulation of \citet{ZuHone2016}
(right). The surface brightness
profiles across these three cases are compared in the bottom panel of
Fig.\ref{Temperatures_finebayedge}, where the cavity toy model is shown to
overestimate the drop in surface brightness across the bay edge, with the discrepancy becoming even worse when an ellipsoidal cavity model is used.  }
            \label{Compare_Perseus_sloshing_cavity_sim}

  \end{center}
\end{figure*}

\begin{figure*}
  \begin{center}
    \leavevmode
\includegraphics[width=\linewidth]{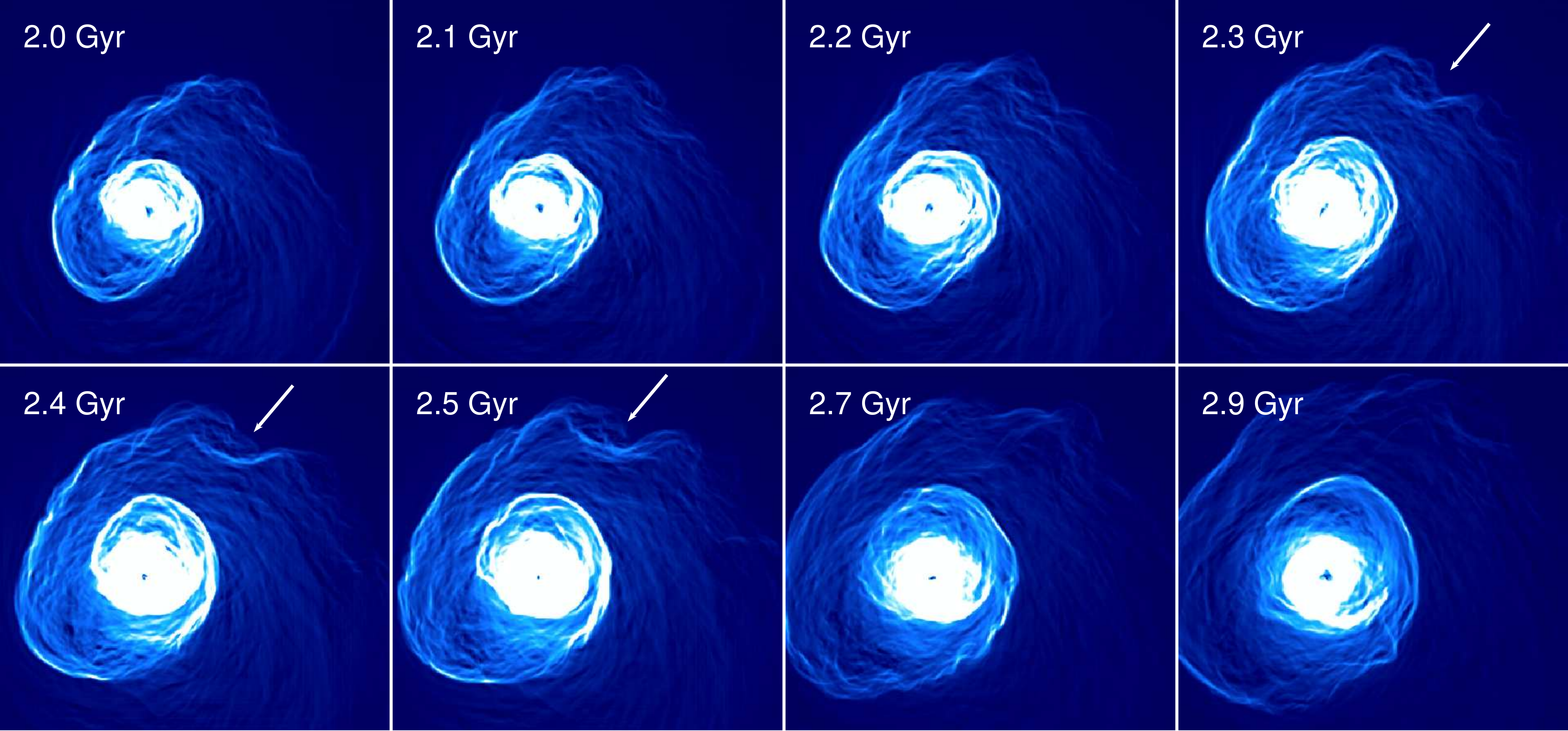}
      \caption{The evolution of KH roll structure in the simulations of Zuhone et
al. 2016 for $\beta= p_{\rm th} / p_{\rm B}$=200, where the time shown is that
which has elapsed since the moment of closest approach betwen the merging clusters. The simulated projected X-ray emissivity images have been filtered using
GGM filtering to emphasise surface brightness edges. We see that KH rolls with
similar structure to the sharp bay seen in Perseus can form (at 2.3 Gyr) and be
destroyed (by 2.7 Gyr) on relatively short time scales of $\sim$0.4Gyr.  }
      \label{GGM_time_ev}
  \end{center}
\end{figure*}

\subsection{Geometry}

One immediately obvious characteristic of the bays is that they are only present on one side of the cluster. Typically, for AGN inflated cavities, 
one would expect there to be a second cavity on the opposite
side of the cluster resulting from the other jet direction. 

The locations of the bays in Perseus and Abell 1795 relative to their main outer cold front are similar, both lying 
around 130 degrees clockwise or counterclockwise from the main cold front. This is shown in Fig. \ref{Compare_Perseus_A1795rot},
in which we have reflected and rotated the image of Abell 1795 to compare to Perseus. This similarity suggests at a link between
the location of the main cold front and the bays in these systems. However the spatial scale of these features in Abell 1795 
is roughly half that of the same features in Perseus, despite both clusters having roughly the same total mass of $7.0\times10^{14}$M$_{\odot}$ (\citealt{Bautz2009}, \citealt{Simionescu2011}). 

The bay is Centaurus differs from those in Perseus and Abell 1795 as it is much closer to the cluster core (around
20kpc from the core compared to 100 kpc in Perseus). In Centaurus there is no significant evidence for a metallicity jump across the bay. There is a high metallicity 
point immediately in front of the bay, though the significance of this is low. The
temperature jump is much less pronounced (a 7 percent jump from 2.7keV to
2.9keV, compared to the $\sim$20 percent jump seen in Perseus and A1795). This
may be because the bay in Centaurus is much closer to the cluster core , where metals
are being deposited into the ICM from the BCG, and where the effect of AGN is greater. Centaurus is well known to have
an extremely high central metal abundance reaching up to 2.5Z$_{\odot}$ (\citealt{Fabian2005}), suggesting that the history of metal 
deposition is more complex than in other clusters.

\section{Cavity scenario}
\label{sec:cavsim}
\subsection{Surface brightness comparison}
Here we investigate the cavity origin scenario for the Perseus bay. We simulated
an image of Perseus in which we took the surface brightness profile from the
region of the cluster on either side of the bay, and extrapolated this outwards.
This acts to `recreate' the original undisturbed ICM surface brightness profile
before the formation of the bay. This surface brightness profile was then
deprojected. 

We then created a 3D cluster toy model, in which each element is weighted by
the X-ray emissivity given by the deprojected surface brightness profile. In the
simulation, a spherical cavity matching the dimensions and location of the bay
was created by setting the X-ray emissivity of the elements within a suitably
sized sphere equal to zero. The X-ray emissivity of the 3D toy model was then
integrated along the line of sight to produce the 2D projected X-ray emissivity
image. Using the Chandra response files and the Chandra PSF, we then produced a
simulated Chandra image for an exposure time matching the real observation, to
which an appropriate background was added. 

The resulting simulated image from the toy model is shown in the central panel of Fig.
\ref{Compare_Perseus_sloshing_cavity_sim}. In the bottom panel of Fig.
\ref{Temperatures_finebayedge} we compare the projected surface brightness
profiles across the real bay (black), and the cavity toy model (green dashed
line). We see that the surface brightness drop expected for a spherically
symmetric cavity is far more severe than we see in Perseus. We repeated this exercise using an ellipsoidal cavity instead, again
matching the radius of curvature of the bay, and with the line of sight depth of the ellipsoid set equal to the width observed in the plane of the sky (60 kpc). 
Increasing the ellipticity of the removed cavity actually further increases the magnitude of the surface brightness drop, increasing the tension
with observations. To match the observed drop in surface brightness, we find that the line of sight depth of the cavity would have to be 
around half its observed width on the sky, so that it is shaped like a rugby ball with the long side being viewed face on. This type of 
unusual geometry is in tension with the idea of rising spherical cap bubbles seen closer in the core of Perseus (\citealt{Fabian2003_filaments}). 
The inner bubbles at around 30-50 kpc from the cluster core already have a distinct spherical cap appearance, which should continue to develop as they rise outwards, 
so cavities at the radial distance of the bay (100 kpc) would be expected to have this form of geometry.

We repeat this exercise with the toy model, but this time for Abell 1795 and Centaurus. When the surface brightness profile across the toy model cavity rim
is compared to the observed profile in the bottom panels of Fig. \ref{Temperatures_finebayedgeA1795andCen}, we again find that
the cavity toy model overestimates the decrement in X-ray surface brightness in both cases.

\subsection{Temperature profile comparison}

Here we investigate the expected temperature profile across the inner rim of a cavity. To achieve this, we use both the deprojected temperature and density profiles
on either side of the bay in our 3D cluster toy model, again `recreating' the original undisturbed ICM. When then removed all of the emission from a spherical region matching
the curvature of the bay, and produced projected spectra across the inner rim, with each temperature component correctly weighted by the emission measures along the line of sight.

For Perseus and Abell 1795, the resultant temperature profiles over the inner rim of the cavity toy model are shown in Figs. \ref{Temperatures_finebayedge} 
and \ref{Temperatures_finebayedgeA1795andCen} as the dashed green lines. In both cases the temperature profile increase is very small, an increase of around 0.15 keV, 
which is much smaller than the observed increase of 0.8 keV for Perseus and 1-1.5keV for Abell 1795. For Centaurus, because the bay is so close to the cluster centre,
it is not possible to accurately use the toy model to reproduce the undisturbed ICM temperature distribution.

Making the line of sight depth of the cavity toy model smaller, in an attempt to match the surface brightness profile, acts to make the temperature jump 
across the inner rim even smaller (since less gas is removed), making the discrepancy with the observed temperature profile even worse. 
We find that for Perseus and Abell 1795, it is not possible to produce a cavity model which matches both 
the surface brightness jump and the temperature jump simultaneously.

\subsection{Comparing with the jet power-metal radius relation}

As a further test of the AGN inflated cavity scenario, we consider the observed relationship between the AGN jet power ($P_{\rm jet}$) and the maximum radius at which an enchancement in metal 
abundance is seen (the Fe radius, $R_{\rm Fe}$). \citet{Kirkpatrick2011} have found a simple power law relationship between these based on observations of clusters
with AGN inflated cavities, which is $R_{\rm Fe}=58 \times P_{\rm jet}^{0.42}$ (kpc), where $P_{\rm jet}$ is in units of $10^{44}$ erg s$^{-1}$. 

The jet power is estimated from the volume of cavity, V, and the pressure of the ICM, P, by dividing the total energy needed to grow the cavity, 4PV, (which is the sum
of the internal energy of the cavity, and the work done in expanding it against the surrounding ICM) by a characteristic timescale over which the cavity has risen, which is typically
taken to be the sound crossing time from the cluster core to the centre of the cavity, $t_{cs}$. Using our best fitting ellipsoidal model for Perseus (which provides a lower bound
to the jet power and the metal radius, for if the cavity were a sphere its volume and thus the jet power would be larger), we find a jet power of 
$1.2\times10^{45}$ erg s$^{-1}$, which gives a metal radius of 165 kpc, much larger than the observed radius of the metal drop of 92 kpc from the cluster core.

Following the same procedure for Abell 1795, we find a jet power of $5\times10^{44}$ erg s$^{-1}$, which gives a metal radius of 114 kpc, again much larger than the observed
radius of the metal drop of 30 kpc from the cluster core. We extended the metallicity profiles outwards for both Perseus and Abell 1795, comparing to the azimuthal average, and 
found no evidence for an enchancement in metal abundance anywhere outwide the bay edges. This lack of a metal abundance excess, and the strong disagreement with the \citet{Kirkpatrick2011}
relation between jet power and metal radius, provides futher evidence against the bays being the inner rims of AGN inflated cavities.

\section{Sloshing simulations}
\label{sec:sloshsim}
\begin{figure*}
  \begin{center}
    \leavevmode
\includegraphics[width=\linewidth]{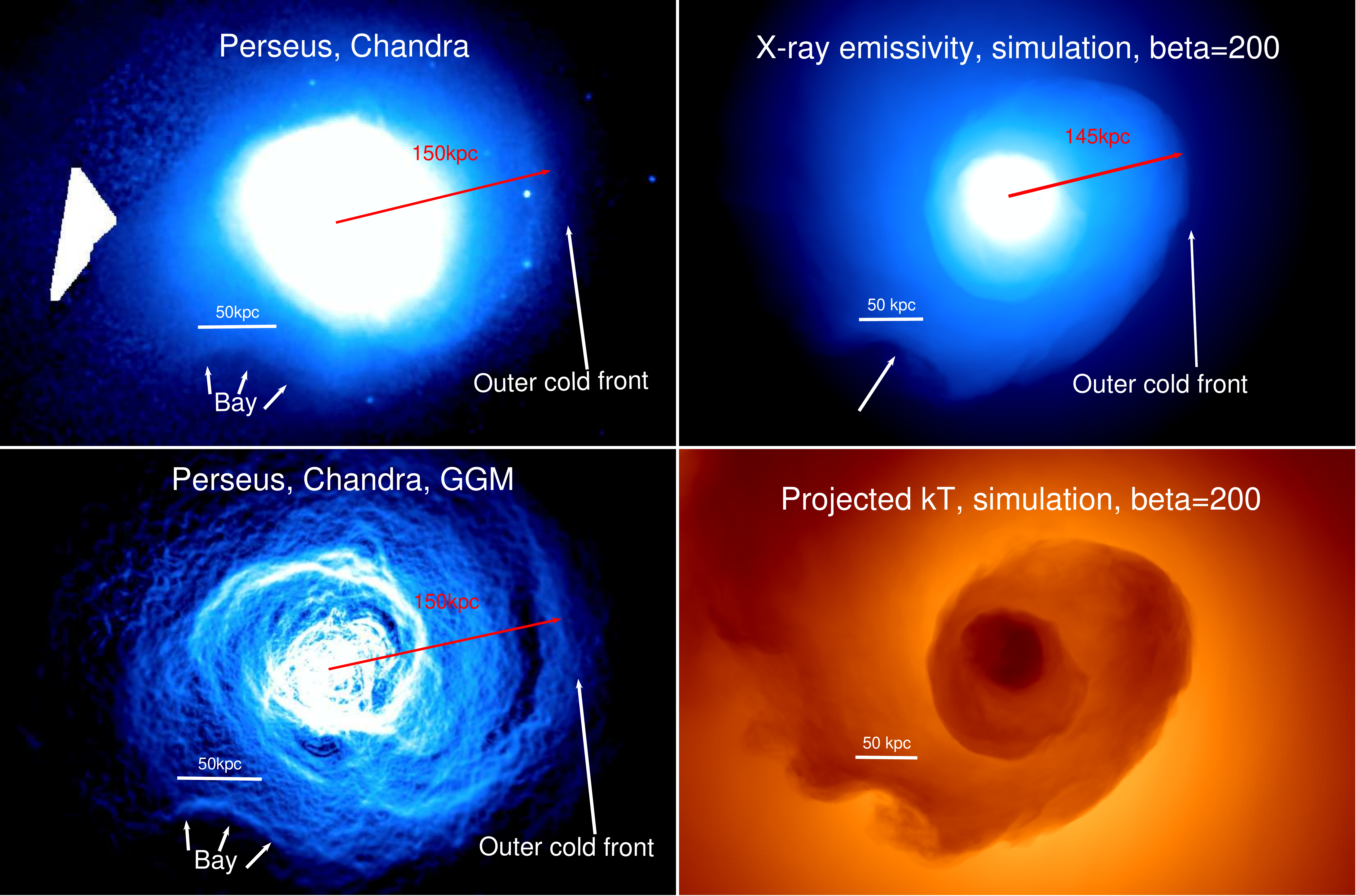}
      \caption{Comparing the outer cold front structure in Perseus with a minor
merger simulation from \citet{ZuHone2016} (`Sloshing of the Magnetized Cool Gas
in the Cores of Galaxy Clusters', $\beta=200$), 2.5 Gyr from the moment of closest approach of the merging clusters. The KH rolls develop into bay-like structures similar to that in the
Perseus observation for a 0.4 Gyr period at this stage. The temperature drop
structure over the KH roll resembles that in Perseus. The brevity of the period in
which these structures are visible (0.4Gyr) before being eroded away may explain
why they are rare. The location of the bay relative to the western outer cold
front is very similar to the location of the KH roll relative to the outer cold
front in the sloshing spiral. The ratio of the sizes of these features is also
very similar between simulation and observation. }
      \label{perseus_comparetobeta200}
  \end{center}
\end{figure*}
To search for concave cold fronts similar to the bay structure in Perseus, we
explored the gas sloshing simulations made publicly available by
\citet{ZuHone2016} in their Galaxy Cluster Merger
Catlog\footnote{http://gcmc.hub.yt/}. Of particular interest is the simulation
'Sloshing of the Magnetized Cool Gas in the Cores of Galaxy Clusters' taken from
\citet{ZuHone2011}, but with higher spatial resolution and an improved treatment
of gravity (see \citealt{Roediger2012b}). In this FLASH AMR simulation, sloshing is
initiated in a massive cluster cool core cluster
($M_{200}$=10$^{15}$M$_{\odot}$) similar to Perseus. The simulations are projected along the axis 
perpendicular to the sloshing direction (i.e. along the z-axis, with sloshing occuring in the x-y plane). We stress that in Perseus, Abell 1795 and Centaurus,
we are unlikely to be viewing the sloshing along such a perfectly perpendicular line of sight, so we expect there to be some line of 
sight projection effects in the real observations.  

As shown in Fig.
\ref{GGM_time_ev}, when the outer cold front rises to around 150kpc from the
core, (similar to the position of the western cold front in Perseus), KH rolls
form (shown by the white arrow), which remain stable over periods of 200-400Myr,
and which have the same X-ray morphology as the Perseus bay. In these simulations, an initially uniform ratio of the thermal pressure ($p_{\rm th}$) to magnetic pressure ($p_{\rm B}$), $\beta= p_{\rm th} / p_{\rm B}$, 
is assumed. As the sloshing progresses, the magnetic field becomes amplified along the cold fronts, restricting transport processes and inhibiting the growth of instabilities. These simulations
have been run for different values of the initial $\beta$ ratio, using the observed range of magnetic field strength from Faraday rotation and synchrotron radiation measurements (1-10$\mu$G) as a guide. Simulation 
runs with $\beta$ = 1000, 500, 200 and 100 are available. We find the best match to the Perseus observations is the $\beta$=200 simulation, which is shown in Fig. \ref{GGM_time_ev} and Fig. \ref{perseus_comparetobeta200}. 

\begin{figure*}
  \begin{center}
    \leavevmode
\includegraphics[width=\linewidth]{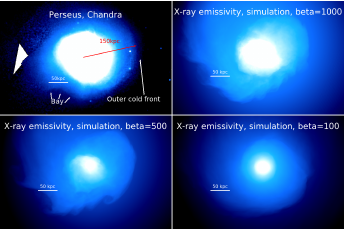}
      \caption{Comparing the Chandra observation of Perseus with simulations of
KH rolls with different values of
$\beta$ (thermal pressure over magnetic pressure) for the same time slice as the
simulation shown in Fig. \ref{perseus_comparetobeta200}. When $\beta$ is high
(1000 and 500), there is much more KH roll structure and the overall morphology
disagrees with the observation. When $\beta$ is low (100), the magnetic field is
strong enough to prevent the formation of large bay shaped KH rolls. }
      \label{perseus_comparebetasims}
  \end{center}
\end{figure*}

\begin{figure*}
  \begin{center}
    \leavevmode
\includegraphics[width=\linewidth]{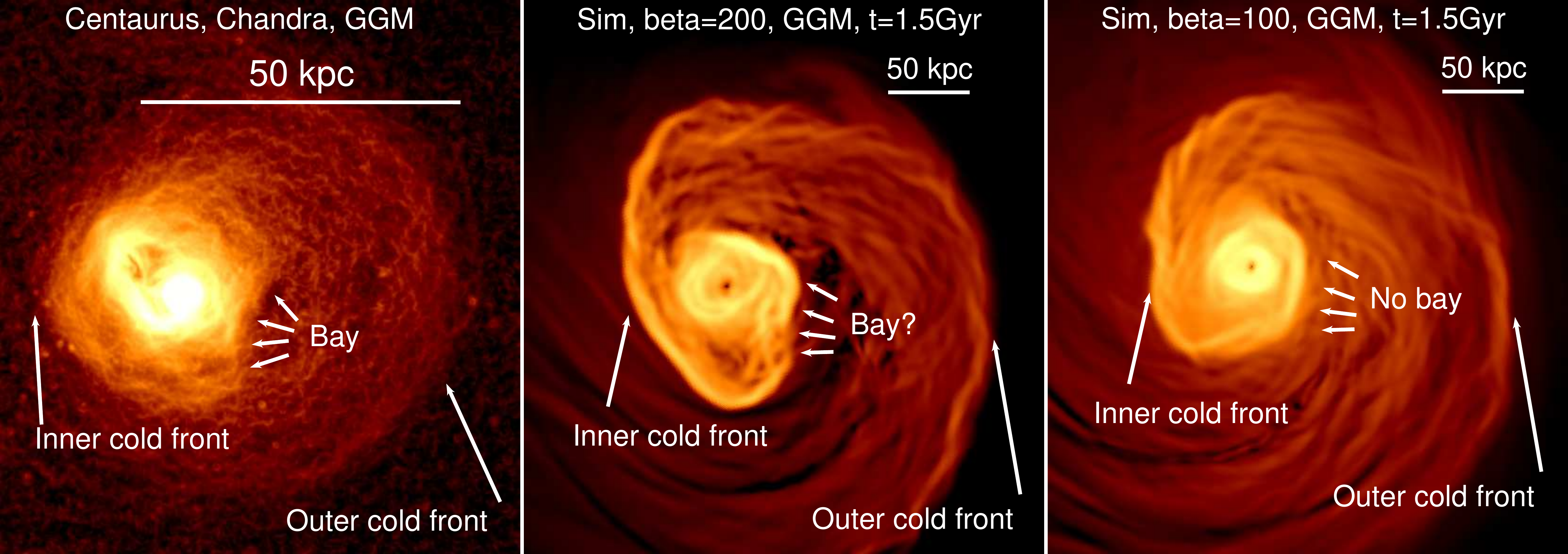}
      \caption{Comparing the bay and cold front morphology in Centaurus (left hand panel) with simulations at 1.5 Gyr from the moment of closest approach of the 
merging clusters. We find 
a similar central bay in the $\beta$=200 X-ray emissivity simulation (middle panel), which is not present in the $\beta$=100 version of the same simulation
with a stronger magnetic field (right hand panel). We note however that the spatial scales on which this bay forms is much larger in the
simulations than in Centaurus. All of these images have been processed with the GGM filter to emphasise gradient structure. }
      \label{Centaurus_comparebetasims}
  \end{center}
\end{figure*}

In the top two panels of Fig. \ref{perseus_comparetobeta200} we compare the
Chandra image of Perseus to the $\beta=$200 sloshing simulation (which we have rotated) at a
stage where the cold fronts are in the same relative location. There is a
striking similarity between the location and size of the bay relative to the
cold front in Perseus and the concave KH roll in the simulation. In both cases the
outer cold front on the right hand side is around 150kpc from the core, and the
bay shaped KH roll forms to the bottom left, around 135 degrees clockwise from the
furthest part of the cold front. In this proposed scenario, the AGN feedback has
destroyed the inner cold front spiral (see the gradient filtered image in the
bottom left of Fig. \ref{perseus_comparetobeta200}), but the outer cold front
with the bay is sufficiently far from the core ($\sim$150 kpc) that it remains
intact.

Using the projected temperature simulation image (bottom right of Fig.
\ref{perseus_comparetobeta200}) we compare the simulated temperature profile
across the bay with the observed one in the top panel of Fig.
\ref{Temperatures_finebayedge}, scaling by a small constant factor to account for the
difference in mass between Perseus ($7\times10^{14}$M$_{\odot}$, \citealt{Simionescu2011}) and the simulated cluster ($10^{15}$M$_{\odot}$). We see that there
is good agreement between the magnitude and shape of the temperature jump. We
then compare the shape surface brightness profile across the bay edge,
shown in the bottom panel of Fig. \ref{Temperatures_finebayedge} as the dashed
blue line. We see that the surface brightness profile is very similar to the
observed profile, and differs significantly from the cavity scenario (green
dashed). The simulations assume a uniform metallicity of 0.3 Z$_{\odot}$, so we are unable to compare the observed metallicity profile jump to a simulated one.

In Fig. \ref{perseus_comparebetasims} we compare the observed cold front
morphology of Perseus with the same time slice of the simulations, but this time
for different values of $\beta$. We see that for higher values of $\beta$ (1000
and 500), the amount of KH roll structure is far greater, and inconsistent with the
observations. For lower values of $\beta$ (100) the formation of KH rolls is
heavily suppressed, and no significant bay-like features form. 

If the bay in Perseus is a KH roll, then this provides the possibility of placing
constraints on the thermal pressure to magnetic pressure ratio by comparing the
observed images with the simulated images. The simulated KH rolls have a large
spatial extent and distinct shape, making them far easier to see than the subtle
structure behind the sharpest part of the cold front. Comparing KH roll structure
between observations and simulations may therefore provide a straightforward way
of constraining $\beta$.

The similar angle between the bay and main cold front in Abell 1795, shown in Fig. \ref{Compare_Perseus_A1795rot}, suggests that
a similar origin may be possible in A1795. However the spatial scale of the features is roughly half that of the Perseus cluster.
It is possible that the reason for this difference is simply that we are seeing Perseus and A1795 at different 
times from the merger that caused the sloshing.

Because the bay in Centaurus is much closer to the core, it is harder to attribute this to sloshing. However, searching 
through the similations, we find a similar morphology of bay and cold front locations in the t=1.5Gyr time frame 
of the $\beta$=200 simulation since the moment of closest approach of the merging clusters, shown in the central panel of Fig. \ref{Centaurus_comparebetasims}. This bay is also not present in the 
same time frame of the higher magnetic field, $\beta$=100 run of the simulations, again showing how senstive such features are to the magnetic pressure level. We stress, however, that the spatial scale of the bay in 
these simulations is around a factor of 4 larger than the bay seen in Centaurus, and that these simulations are for a much more massive cluster ($10^{15}$M$_{\odot}$) 
than the Centaurus cluster ($2\times10^{14}$M$_{\odot}$, \citealt{Walker2013_Centaurus}), so the similarity is purely qualitative in this case. We note 
that the simulation images show linear features running parallel to and behind the outer main cold front, similar to the linear features found in the
Chandra data in \citet{Sanders2016} (their figure 7). As discussed in \citet{Werner2016}, which found similar features in the Virgo cluster cold front, 
in the simulations these linear features are brought about by alternating areas of weak and strong magnetic pressure.

\section{Conclusions}
\label{sec:conclusions}
We have investigated the origin of concave `bay' shaped structures in three
nearby clusters with deep Chandra observations (Perseus, Centaurus and Abell
1795), which have in the past been interpreted as the inner rims of cavitities
inflated by AGN feedback. All three bays show temperature jumps coincident with
the surface brightness jump, and have widths of the order of the Coulomb mean free
path, making them consistent with cold fronts but for the fact that they concave instead of a convex. 

We find that the observed temperature, density, metal abundance and radio distributions around the bays are incompatible with a cavity origin. By comparing with simulations
of gas sloshing from \citet{ZuHone2016}, we find that the observed properties of the bays are consistent with large Kelvin Helmholtz rolls, which produce similar concave cold front structures. 

To test whether these bays could be cold fronts, we explored simulations of gas
sloshing in a massive cluster from \citet{ZuHone2016} to search for similar
features. We find that, when the sloshing has developed to the point seen in
Perseus, with an outer cold front at around 150 kpc from the core, large bay
shaped KH rolls resulting from Kelvin Helmholtz instabilities can form. The
relative size and location of these KH rolls to the cold front structure bears a
striking similarity to the position of the bay in Perseus. The profiles of
temperature and surface brightness are also in good agreement with the observed
profile across the Perseus bay. We also find that the central bay in the Centaurus cluster, which lies much closer to the cluster core (around 15kpc) than those in Perseus and A1795, can qualitatively be explained 
by gas sloshing rather than AGN feedback. 

The shape of these instabilities is sensitive to the ratio of the
thermal pressure to magnetic pressure, $\beta= p_{\rm th} / p_{\rm B}$. We find
the best match to simulations with $\beta=200$. When $\beta$ is higher than this
(1000 and 500) the level of instabilities is too great compared to observations,
while for the $\beta$=100 simulations the magnetic field strength is strong
enough to prevent the formation of the instabilities. Due to the size of the
KH rolls, they are far easier to see than the subtle differences in width of the
cold fronts at their sharpest points. If the bay in Perseus is a Kelvin-Helmholtz roll, then
this may provide a straightforward way of constraining $\beta$ by comparing the
observed image with simulated X-ray images. In particular, it potentially provides a simple way of ruling out simulations where the magnetic field is too high, as these
prevent the formation of KH rolls, an effect which is far easier to see than subtle changes in the width of the traditional main convex cold front.

Whilst it is possible to put order of magnitude constraints on the magnetic field, more precise constraints using this method are likely challenging due to the sensitivity of KH instabilties to a number of factors. The initial 
perturbations used in the simulations are a factor, in that stronger initial perturbations may lead to more pronounced KH rolls. The clarity of KH rolls at late stages of the simulations is also dependent on the spectrum 
of the initial perturbations. 

One potential advantage of constraining the initial magnetic field using cold front structure is that we are probing the average magnetic field throughout the whole volume of the cluster, since the cold fronts rise from the 
core outwards and sample large volumes of the ICM. Measurements of the magnetic field in clusters using the Faraday rotation measure (RM) are typically limited to small sight lines  
probing small sections of the cluster ICM (see \citealt{Taylor2006}), and so can vary considerably by an order of magnitude (typically 1-10$\mu$G) due to fluctuations in the magnetic field on scales of 5-10 kpc 
(\citealt{Carilli2002review}). Comparing the observations with sloshing simulations therefore provide unique, large scale average constraints on the overall magnetic field throughout the cluster volume.

In this paper we have focussed on three nearby clusters which have deep Chandra and radio data. A more systematic 
future study will be required, combining X-ray and radio data, to 
determine how common the `bay' features are in the cluster population as a whole. Due to the complex spiral structure of gas sloshing, projection effects are more severe than for a situation consisting of 
AGN inflated cavities in 
an otherwise undisturbed ICM. The best candidates for finding bays in other sloshing cluster cores are systems where we are viewing along a line of sight that is close to perpendicular to the plane of the sloshing.

The simulations we have looked at only vary the cluster magnetic pressure level whilst keeping the viscosity constant. 
The ICM viscosity also significantly affects the development of KHI rolls (\citealt{Roediger2013}), with a greater viscosity inhibiting the formation
of instabilities. Future work is necessary to understand the relative contributions of these two factors to the overall cold front structure (\citealt{Zuhone2016review}).

\section*{Acknowledgements}
We thank the referee, E. Roediger, for helpful suggestions which improved the paper.
SAW was supported by an appointment to the NASA Postdoctoral Program at the
Goddard Space Flight Center, administered by the Universities Space Research
Association through a contract with NASA.
JHL is supported by NSERC through the discovery grant and Canada Research Chair programs, as well as FRQNT. 
ACF acknowledges support from ERC
Advanced
Grant FEEDBACK. We thank John ZuHone for making the simulations shown in this paper publicly available. This
work is based on observations obtained with the \emph{Chandra} observatory, a 
NASA mission.
\bibliographystyle{mn2e}
\bibliography{Bays_v3}

\appendix
\section[]{Observations and spectral extraction regions}
\label{appendix_obs}

\begin{table*}
\begin{center}
\caption{Chandra data used in this paper.}
\label{obsdata}
\leavevmode
\begin{tabular}{l| l|l | l | l | l} \hline \hline
Object & Obs ID&Exposure (ks)&RA&Dec&Start Date\\ \hline
Perseus & 3209&95.77&03 19 47.60&+41 30 37.00&2002-08-08\\
& 4289&95.41&03 19 47.60&+41 30 37.00&2002-08-10\\
& 4946&23.66&03 19 48.20&+41 30 42.20&2004-10-06\\
& 4947&29.79&03 19 48.20&+41 30 42.20&2004-10-11\\
& 6139&56.43&03 19 48.20&+41 30 42.20&2004-10-04\\
& 6145&85.00&03 19 48.20&+41 30 42.20&2004-10-19\\
& 4948&118.61&03 19 48.20&+41 30 42.20&2004-10-09\\ 
& 4949&29.38&03 19 48.20&+41 30 42.20&2004-10-12\\
& 6146&47.13&03 19 48.20&+41 30 42.20&2004-10-20\\
& 4950&96.92&03 19 48.20&+41 30 42.20&2004-10-12\\
& 4951&96.12&03 19 48.20&+41 30 42.20&2004-10-17\\
& 4952&164.24&03 19 48.20&+41 30 42.20&2004-10-14\\
& 4953&30.08&03 19 48.20&+41 30 42.20&2004-10-18\\ \hline
Centaurus & 504&31.75&12 48 48.70&-41 18 44.00&2000-05-22\\
& 4954&89.05&12 48 48.90&-41 18 44.40&2004-04-01\\
& 4955&44.68&12 48 48.90&-41 18 44.40&2004-04-02\\
& 5310&49.33&12 48 48.90&-41 18 44.40&2004-04-04\\
& 16223 &	180.0 	&12 48 48.90 &	-41 18 43.80 &	2014-05-26 \\
& 16224 &	 	42.29 &	12 48 48.90 	&-41 18 43.80 	& 	2014-04-09 \\ 	
& 16225 &	 	30.1 	&12 48 48.90 &	-41 18 43.80 	& 	 	2014-04-26 \\
& 16534 &	 	55.44 &	12 48 48.90 &	-41 18 43.80 	 &	 	2014-06-05 \\
& 16607 &	 	45.67 &	12 48 48.90 &	-41 18 43.80 	 &	 	2014-04-12 \\
& 16608 &	 	34.11 &	12 48 48.90 &	-41 18 43.80 	 &	 	2014-04-07 \\
& 16609 &	 	82.33 &	12 48 48.90 &	-41 18 43.80 	 &	 	2014-05-04 \\ 	
& 16610 &	 	17.34 &	12 48 48.90 &	-41 18 43.80 	 &	 	2014-04-27\\ \hline
Abell 1795 & See tables A1 and A2 from \citet{Walker2014_A1795} &&&&\\ \hline

\end{tabular}
\end{center}
\end{table*}

\begin{figure*}
  \begin{center}
    \leavevmode
\includegraphics[width=\linewidth]{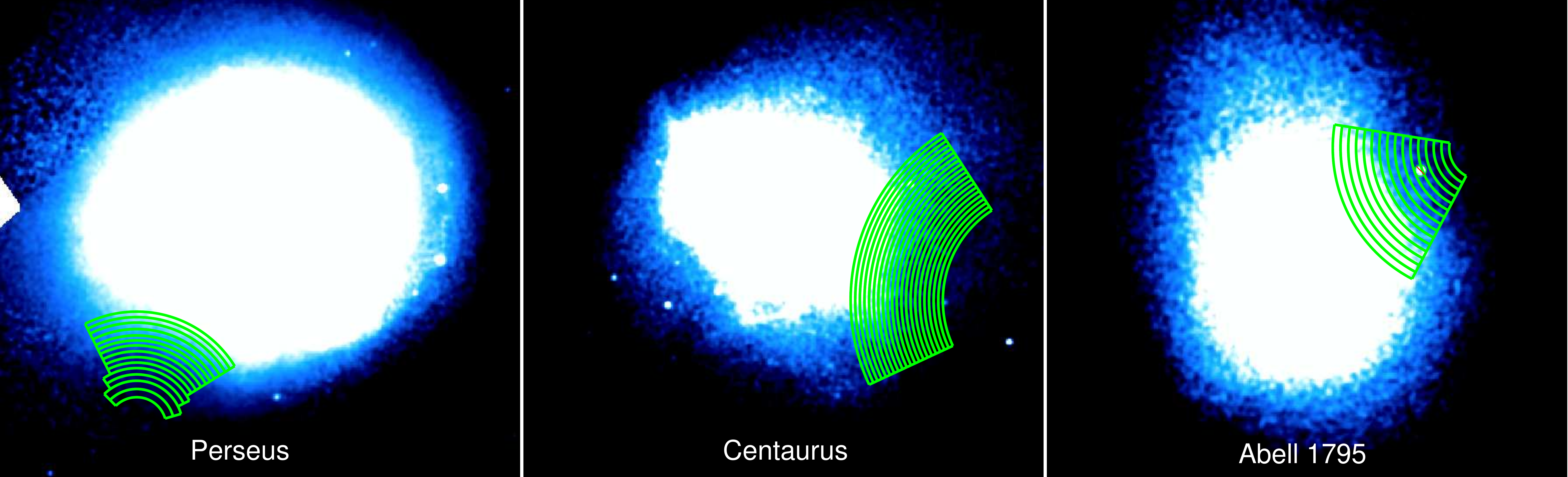}
      \caption{The regions used to extract the spectra for the profiles of temperature and metal abundance shown in Figs. \ref{Temperatures_finebayedge} and \ref{Temperatures_finebayedgeA1795andCen}, overlaid on 
the Chandra observations. }
      \label{Cluster_profile_regions}
  \end{center}
\end{figure*}

\end{document}